# Event Soliton Formation in Mixed Energy-Momentum Gaps of Nonlinear Spacetime Crystals

Liang Zhang[1], Zhiwei Fan[2*], Yiming Pan[1*]

[1]School of Physical Science and Technology and Center for Transformative Science, ShanghaiTech University, Shanghai 200031, China

[2]School of Engineering, Newcastle University, Newcastle upon Tyne NE1 7RU, UK

**Abstract**

We report the formation of a novel soliton—termed "event soliton"—in nonlinear photonic spacetime crystals (STCs). In these media, simultaneous spatiotemporal periodic modulation of the dielectric constant generates mixed frequency ($\omega$) and wavevector ($k$) gaps. Under Kerr nonlinearity, the event solitons emerge as fully localized entities in both spacetime and energy-momentum domains, providing a tangible demonstration of the concept of 'event' in relativity. The $\omega k$-gap mixture arises from the coexistence and competition between time-reflected and Bragg-reflected waves due to the spatiotemporal modulation. We propose a new partial differential equation to capture various spatiotemporal patterns and present numerical simulations to validate our theoretical predictions, reflecting a three-way balance among $k$-gap opening, $\omega$-gap opening, and nonlinearity. Our work opens avenues for fundamental studies and fosters experimental prospects for implementing spacetime crystals in both time-varying photonics and periodically driven condensed matter systems.

Recently, the study of spacetime crystals (STCs) in photonics — also known as spatiotemporal photonic crystals [1] or spacetime metamaterials [2,3] — has garnered widespread interest. Unlike photonic crystals (PhCs) or photonic time crystals (PTCs) [4–14], photonic STCs feature a dielectric constant that varies periodically in both space and time, as described by $\epsilon(x,t) = \epsilon(x + \Lambda, t + T)$, where $\Lambda$ and $T$ are the spatial and temporal periodicities, respectively. In PhCs, spatial periodicity leads to an energy gap ($\omega$-gap) via Bragg reflection [15], while temporal periodicity in PTCs results in a momentum gap ($k$-gap) through time reflection [7,16,17]. STCs support both $\omega$-gaps and $k$-gaps, enabling distinctive bandgap engineering. Y. Sharabi et al. introduced the concept of a "mixed energy-momentum gap" ($\omega k$-gap) [1,18], which emerges when these two gaps converge and overlap, thereby yielding intricate $\omega k$-gapped mode dynamics.

In a dispersive medium, wave propagation exhibits frequency-dependent wavevector. For a STC, when the dispersion relation contains a gap (forbidden band), it is more precisely described as a dispersion-dissipation relation [19], as both the frequency and wavevector acquire complex values: the real parts govern dispersion, while the imaginary components represent dissipation. Notably, modes within $\omega$-gap or $k$-gap exhibit distinct dissipative behaviors based on the sign of the imaginary parts. For $\omega$-gapped modes, the momentum (or wavevector) is complex, with the imaginary part scaled by the gap size, leading to spatial attenuation under energy conservation, with the unphysical amplifying mode excluded. Conversely, $k$-gap modes feature complex frequencies whose imaginary parts correspond to the gap size, resulting in exponential growth or decay over time [14,20]. Bear in mind that this growing mode is physical, in the sense that the system extracts energy from temporal modulation, highlighting a distinction between $k$-gapped and $\omega$-gapped modes. This property is pivotal for lasing and amplification in PTCs [21,22]. In a STC, both the $k$-gap and $\omega$-gap coexist, enabling their controlled convergence through spatiotemporal modulation to merge into a mixed gap (see Fig.1h). Within the mixed gap, the systems exhibit spatial and temporal amplification or attenuation simultaneously due to both complex frequencies and wavevectors, thus dictating unstable propagation of $\omega k$-gapped modes.

A key question in this work is whether nonlinear mechanisms can stabilize the mode dissipation within $\omega k$-gaps, a possibility that has not yet been reported. In Bragg gratings, nonlinearity generates localized states known as Bragg solitons, which are spatially confined $\omega$-gapped solutions that remain stationary. These gap solitons were extensively studied in the late 1980s [23–30] and can be well characterized by the nonlinear Schrödinger equation (NLSE) in the form of $[(1 - \partial_{xx})\psi - 2\psi^3 = 0]$ [31]. Similarly, nonlinearity in the $k$-gap region gives rise to temporally localized modes.

Notably, the $k$-gap soliton reported by Y. Pan et al. [32] is spatially homogeneous and propagates at a superluminal velocity while preserving Einstein causality, and is accurately modelled by the NLSE as $[(1 - \partial_{tt})\psi - 2\psi^3 = 0]$. This $k$-gap soliton exhibits a dual spacetime structure in comparison to the Bragg soliton achieved by substituting the spatial derivative ($\partial_{xx}$) with temporal derivative ($\partial_{tt}$). These gap solitons are compared schematically in Fig. 1.

Here, we hypothesize the presence of an "event soliton" within the mixed $\omega k$-gap, exhibiting both spatial and temporal localization, as depicted in Fig.1l and 1h. To facilitate such a localized solution, without mentioning the specific setting, we intuitively propose a normalized nonlinear wave equation: $(1 - \partial_{tt})(1 - \partial_{xx})\psi - 4\psi^3 = 0$. We investigate how Kerr nonlinearity ($\psi^3$) enables four unstable $\omega k$-gapped modes to mutually counterbalance, forming an event soliton. This event-like localization can be interpreted as a nonlinear mixing among the gapped modes, analogous to $\text{sech } x \text{ sech } t = 4/(e^{x+t} + e^{-x+t} + e^{-x-t} + e^{x-t})$, displaying temporal spontaneity and spatial uncertainty. Our hypothesis reveals a nontrivial spatiotemporally localized solution in our setting of a nonlinear STC, where energy and momentum are pinned in the mixed $\omega k$-gap, and its dynamics satisfies the newly-constructed nonlinear wave equation. Finally, we discuss the implications of our findings for the experimental realization of nonlinear STCs.

**Modelling**. To derive the event soliton and its governing equation as shown in Figs. 1h and 1l, we start by modelling the nonlinear STC depicted in Fig. 1d. In a non-magnetic dielectric medium without free charges and moving currents ($\rho = 0, \boldsymbol{J} = 0$), Maxwell's equations in a source-free medium are given by $\nabla \times \boldsymbol{E} = -\frac{\partial \boldsymbol{B}}{\partial t}, \nabla \times \boldsymbol{B} = \mu_0 \frac{\partial \boldsymbol{D}}{\partial t}$, with divergence equations $\nabla \cdot \boldsymbol{D} = 0, \nabla \cdot \boldsymbol{B} = 0$, where $\boldsymbol{E}$ is the electric field, $\boldsymbol{D}$ is the electric displacement, $\boldsymbol{B}$ is the magnetic field, and $\mu_0$ is the vacuum permeability. Applying the curl operator ($\nabla \times$) to the first equation and substituting the result into the second yields:

$$\frac{\partial^2 \boldsymbol{D}}{\partial t^2} = -\frac{1}{\mu_0} \nabla \times (\nabla \times \boldsymbol{E}). \tag{1}$$

Assuming a centrosymmetric medium where second-order nonlinearity is negligible, the electric displacement field can be expressed as: $D = \epsilon(x,t)E = \epsilon_0(\epsilon_1(x,t) + \chi_3|E|^2)E$, where the Kerr nonlinearity $\chi_3$ is dominant. The linear dielectric constant, $\epsilon_1$, is modulated periodically in space and time and is given by $\epsilon_1(x,t) = \epsilon_r \tilde{\epsilon}(x)\tilde{\epsilon}(t) = \epsilon_r(1 + \delta_1 cos(\Omega t))(1 + \delta_2 cos(Gx))$. Here, $\epsilon_r$ is the mean permittivity, $\delta_{1,2} < 1$ as small modulation strengths, and $\Omega = 2\pi/T$, $G = 2\pi/\Lambda$ specify the temporal and spatial modulation frequencies, with periods $T$ and $\Lambda$, respectively. In a one-dimensional isotropic medium, the electric field can be reciprocally expressed in terms of the displacement as: $E = \frac{D}{\epsilon_0 \epsilon_1} - \frac{\chi_3 D^3}{\epsilon_0^2 \epsilon_1^4}$, where $\chi_3/\epsilon_0 \epsilon_1^4$ denotes the reciprocal third-order nonlinearity [33]. Substituting this into Eq. 1 yields

$$\frac{\partial^2 D}{\partial t^2} = \frac{1}{\mu_0} \frac{\partial^2}{\partial x^2} \left( \frac{D}{\epsilon_0 \epsilon_1} - \frac{\chi_3 D^3}{\epsilon_0^2 \epsilon_1^4} \right). \tag{2}$$

In time-varying media, the electric displacement D is favored over the electric field E because D remains continuous across temporal boundaries, as dictated by Gauss's law ($\nabla \cdot \mathbf{D} = 0$), whereas E may be discontinuous [37]. These temporal boundaries give rise to phenomena such as time reflection and time refraction [17]. For small modulation amplitudes $\delta_{1,2} \ll 1$ and weak nonlinearity $\chi_3$, Eq. 2 can be simplified to:

$$\frac{(1 + \delta_2 \cos Gx)}{c^2} \frac{\partial^2 \tilde{E}}{\partial t^2} = (1 - \delta_1 \cos \Omega t) \frac{\partial^2 \tilde{E}}{\partial x^2} - \beta |\tilde{E}|^2 \frac{\partial^2 \tilde{E}}{\partial x^2}, \tag{3}$$

where $\tilde{E} = D/\tilde{\epsilon}(x)$ is the reduced electric displacement and $c = c_0/\sqrt{\epsilon_r} = c_0/n_0$ is the speed of light in the medium, with $n_0 = \sqrt{\epsilon_r}$ being the refractive index and $c_0$ the light speed in vacuum. We approximate $\tilde{\epsilon}^{-1}(t) = 1 - \delta_1 \cos \Omega t$, and focus on the self-phase-modulation term in the nonlinearity by redefining the Kerr coefficient $\beta = 3\chi_3/\epsilon_0 \epsilon_r^3$. Equation (3) encapsulates the modulated electromagnetic waves in the nonlinear STC and serves as the base model for our analysis.

To derive the event soliton from Eq. 3, we further simplify it to nonlinear coupled-mode equations [34], based on the Floquet-Bloch theorem. To seek Floquet-Bloch waves as the sum of spatiotemporally modulated forward and backward waves, we assume:

$$\tilde{E}(x,t) = A_f e^{i\frac{G}{2}x - i\frac{\Omega}{2}t} + A_b e^{-i\frac{G}{2}x - i\frac{\Omega}{2}t} + A_f^* e^{-i\frac{G}{2}x + i\frac{\Omega}{2}t} + A_b^* e^{i\frac{G}{2}x + i\frac{\Omega}{2}t}, \tag{4}$$

where $A_f$, $A_b$ are the complex amplitudes of the forward- and backward-propagating waves, and their conjugates $A_f^*$, $A_b^*$. Substituting the ansatz into Eq. 3 and applying the slowly-varying envelope approximation ($\delta_{1,2} \ll 1$, $\beta \ll 1$) yields the coupled-mode equations:

$$iG \frac{\partial A_f}{\partial x} + i \frac{\Omega}{c^2} \frac{\partial A_f}{\partial t} + \lambda A_f + \kappa_1 A_b^* + \kappa_2 A_b = \gamma \left( |A_f|^2 + 2|A_b|^2 \right) A_f,$$

$$-iG \frac{\partial A_b}{\partial x} + i \frac{\Omega}{c^2} \frac{\partial A_b}{\partial t} + \lambda A_b + \kappa_1 A_f^* + \kappa_2 A_f = \gamma \left( 2|A_f|^2 + |A_b|^2 \right) A_b,$$

$$-iG\frac{\partial A_f^*}{\partial x} - i\frac{\Omega}{c^2}\frac{\partial A_f^*}{\partial t} + \lambda A_f^* + \kappa_1 A_b + \kappa_2 A_b^* = \gamma(|A_f|^2 + 2|A_b|^2)A_f^*,$$

$$iG\frac{\partial A_b^*}{\partial x} - i\frac{\Omega}{c^2}\frac{\partial A_b^*}{\partial t} + \lambda A_b^* + \kappa_1 A_f + \kappa_2 A_f^* = \gamma(2|A_f|^2 + |A_b|^2)A_b^*. \quad (5)$$

Here, $\lambda = (\Omega^2/c^2 - G^2)/4$ represents the mismatch between temporal and spatial modulation frequencies. The coupling strengths are given by $\kappa_1 = \delta_1 G^2/8$, $\kappa_2 = \delta_2 \Omega^2/8c^2$, and the nonlinear coefficient is $\gamma = -3\beta G^2/4$. These equations capture the scattering between Floquet-Bloch waves, including Bragg reflection ($A_f \leftrightarrow A_b$), and time reflection ($A_f \leftrightarrow A_b^*$). The decomposition into forward and backward propagating waves reveals self-phase modulation and cross-phase modulation terms arising from the nonlinear mixing.

**Mixing the energy and momentum gaps**. To derive the $\omega k$-gapped dispersion, we rewrite the Floquet-Bloch waves as a spinor with four components $\psi = (A_f, A_b, A_f^*, A_b^*)^T$. Equation (5) can be recast in a compact form

$$\left(\left(i\frac{\partial}{\partial x}\right)\sigma_z\tau_z + \frac{\Omega}{c^2}\left(i\frac{\partial}{\partial t}\right)\sigma_z\tau_0 + \lambda\sigma_0\tau_0 + \kappa_1\sigma_x\tau_x + \kappa_2\sigma_0\tau_x - \gamma\mathcal{L}_{\text{NL}}[\psi]\right)\psi = 0, \quad (6)$$

where $\sigma_i, \tau_j$ are Pauli operators, with $\sigma_i\tau_j = \sigma_i \otimes \tau_j$ for $i,j = 0, x, y, z$. The nonlinear term $\mathcal{L}_{\text{NL}}[\psi] = |A_f|^2\sigma_0(3\tau_0 - \tau_z)/2 + |A_b|^2\sigma_0(3\tau_0 + \tau_z)/2$ accounts for both self-phase and cross-phase modulations. To obtain the band structure associated with the linear part of Eq. 6, we consider plane-wave solutions of the form $\psi = \chi e^{iPx - iEt}$, where $\chi$ is a spinor with $P$ being the effective wavevector and $E$ the effective frequency. By suppressing the nonlinear term ($\gamma = 0$), the resulting band structure for the coupled modes is

$$\left((PG)^2 - \left(\frac{E\Omega}{c_r^2}\right)^2 - \lambda^2 - \kappa_1^2 + \kappa_2^2\right)^2 = 4\left(\frac{E\Omega\lambda}{c_r^2}\right)^2 + 4(\lambda\kappa_1)^2, \quad (7)$$

Figure 2 illustrates the band structure engineering of the forward propagating wave in the STCs. Details are provided in supplementary Section 2 of the SM file. We define the modulation ratio $r = \Omega/Gc$ as a control parameter, so that the mismatch $\lambda = G^2(r^2 - 1)/4$. Without loss of generality, we hold G constant and set the same modulation amplitudes $\delta_1 = \delta_2 = \delta$ (notice that still $\kappa_1 \neq \kappa_2$). By altering $\Omega$ (corresponding to r), the band structure can be engineered to maimtain the fixed energy $\omega$-gap while bringing the momentum $k$-gap closer to it. For $r \ll 1$ (Fig. 2a), the $\omega$-gap and $k$-gap remain well-separated. The complex wavevectors are confined within the $\omega$-gap, and the complex frequencies reside in the $k$-gap, appearing in pairs. At the bandgap edges, the group velocity of the $\omega$-gap is zero, whereas for the $k$-gap, it is infinite.

As the ratio r increases to 0.85 (Fig.2b), the $k$-gap draws critically closer to the $\omega$-gap. By r=0.95, the two gaps merge into a single mixed gap (the so-called $\omega k$-gap), where both frequency and wavevector become simultaneously complex. At this point, the lower branch band edge exhibits infinite group velocity $v_g^{(-)} \to \infty$, and the upper branch has zero group velocity $v_g^{(+)} \to 0$. When r slightly surpasses 1, the mixed $\omega k$-gap persists, but the group velocities at the band edges interchange, resulting in $v_g^{(-)} \to 0$ and $v_g^{(+)} \to \infty$. Within the $\omega k$-gap, four gapped modes emerge, displaying simultaneous amplification or attenuation in spacetime, leading to complex unstable patterns. Increasing r further, for example to 1.2 (Fig. 2e) or 1.5 (Fig. 2f), leads to the re-separation of the $k$-gap from $\omega$-gap.

In the bandgap engineering, an intriguing phenomenon occurs at r=1 (Fig. 2g), termed "light-cone modulation". At this ratio, these $\omega k$-gapped modes converge such that their real parts collapse to the modulation center $(G/2, \Omega/2)$ in the band. While, the imaginary parts of the frequency and wavevector emerge to build up a dissipation relation around the center, given by $Im(\omega)^2 = Im(k)^2$, as shown in Fig. 2h. Notably, the plane-wave solution $\psi = \chi e^{iPx - iEt}$ becomes invalid for describing these gapped modes. Instead, we introduce a dissipative ansatz $\psi' = \chi e^{\xi x - \eta t}$, where $\eta, \xi$ are assumed to be real, corresponding to the imaginary parts of the frequency and wavevector. At the center (r=1), the linear dissipation relation $|\eta| = c|\xi|$ holds (see Fig. S2 in the SM file). As shown in Fig. 2i, four degenerate dissipative modes emerge: $\chi_m e^{\eta t - \xi x}, \chi_m e^{\eta t + \xi x}, \chi_m e^{-\eta t + \xi x}$, and $\chi_m e^{-\eta t - \xi x}$, where $m = a, b, c, d$ denote the mode indices. These dissipative modes determine the possible spatiotemporal behavior and pattern formation within the $\omega k$-gaps when the nonlinearity ($\mathcal{L}_{\mathrm{NL}}[\psi]$) is considered.

**Soliton formation in the mixed $\omega k$-gap.** Under light-cone modulation, the gapped eigenmode spinors are $\chi_a = (-i, i, -i, i)^T/2, \chi_b = (i, -i, -i, i)^T/2, \chi_c = (-1, -1, 1, 1)^T/2, \chi_d = (1, 1, 1, 1)^T/2$, respectively. In our analysis, we seek spatiotemporally localized solutions of the form $\psi = a(x,t)\chi e^{iP_0 x - iE_0 t}$ where $P_0^2 = E_0^2/c^2$, and at the modulation center $P_0 = 0, E_0 = 0$. By selecting one of these spinors, $\chi_d$, and setting λ=0 and $\kappa_1 = \kappa_2 = \kappa$, we substitute the ansatz into Eq. 6, yielding a differential equation for the slowly varying amplitude $a(x,t)$, given by

$$\left(1 - \frac{\Omega^2}{4\kappa^2 c_r^4}\frac{\partial^2}{\partial t^2}\right)\left(1 - \frac{G^2}{4\kappa^2}\frac{\partial^2}{\partial x^2}\right)a - \frac{3\gamma}{8\kappa}|a|^2 a = 0, \tag{8}$$

which admits an analytical solution in the $\omega k$-gap

$$a = \frac{4\sqrt{2\kappa}}{\sqrt{3\gamma}}\text{sech}\left(\frac{2\kappa c^2}{\Omega}t\right)\text{sech}\left(\frac{2\kappa}{G}x\right). \tag{9}$$

Accordingly, the electric displacement D is then given by

$$D(x,t) = \frac{8\sqrt{2\kappa}}{\sqrt{3\gamma}}(1 + \delta_2 \cos Gx)\text{sech}\left(\frac{2\kappa c^2 t}{\Omega}\right)\text{sech}\left(\frac{2\kappa x}{G}\right)\cos\left(\frac{Gx}{2}\right)\cos\left(\frac{\Omega t}{2}\right). \tag{10}$$

The detailed derivation of this equation from $\chi_d$ can be found in the SM Section 4. Eq. 10 demonstrates a dual-localized structure in time and space arising from a delicate nonlinear mixing of the four $\omega k$-gapped modes. This nonlinear mixing aligns with strategies observed in Bragg soliton and $k$-gap soliton formation, extending similar principles into the spacetime domain. Full derivations are provided in the SM file.

This event soliton (Eq. 9) is one of our key achievements in this work, with the following features: the peak intensity scales proportionally with the modulation strength ($\kappa$) and is inversely proportional to the nonlinearity ($\gamma$), oscillating with a frequency half of the temporal modulation and a wavevector half of the spatial modulation. The $\omega k$-gap soliton is called an event soliton due to simultaneous localization in spatiotemporal and energy-momentum domains, analogous to a classical event in special relativity. The wavepacket size of event soliton is determined by the $\omega k$-gap size, indicating a constraint imposed by the uncertainty principle.

**Generating spatiotemporal patterns from a localized input.** To numerically explore the general behavior, we simplify the nonlinear wave equation (Eq. 8) to a normalized partial differential equation:

$$(1 - \partial_{tt})(1 - \partial_{xx})\psi - 4\psi^3 = 0, \tag{11}$$

where time t, space x and field $\psi$ are dimensionless variables renormalized from Eq. 8. A direct check confirms the normal form of event soliton: $\psi = \text{sech}\, x\, \text{sech}\, t$. Under specific constraints, Eq. 11 can further yield $k$-gap solitons or Bragg solitons. For instance, assuming spatial independence $\partial_{xx}\psi = 0$, Eq. 11 reduces to $(1 - \partial_{tt})\psi -$ $4\psi^3 = 0$, which admits a $k$-gap soliton solution $\psi = \frac{1}{\sqrt{2}}\text{sech}\, t$. On the other hand, when the solution becomes temporally independent $\partial_{tt}\psi = 0$, Eq. 11 simplifies to $(1 - \partial_{xx})\psi - 4\psi^3 = 0$, yielding the Bragg soliton solution $\psi = \frac{1}{\sqrt{2}}\text{sech}\, x$. Additionally, in the absence of the nonlinear term when $(1 - \partial_{tt})(1 - \partial_{xx})\psi = 0$, we obtain analytical solutions $\psi = e^{\pm x \pm t}, e^{\pm x \mp t}$, corresponding to the four $\omega k$-gapped modes (Fig. 2i).

We systematically studied Equation (11) that governs event solitons in spacetime crystals. We found that different spatiotemporal patterns emerge depending on initial conditions. For a low-amplitude plane wave with a homogeneous ($\psi(x, 0) = 0.0001$) is initialized, exponential amplification occurs due to the $\omega k$-gap, which eventually triggers nonlinearity and forms a $k$-gap soliton. As shown in Fig. 3a, this soliton remains finite in time but extends as a plane wave in space, reflecting a two-way balance between nonlinearity and $k$-gap opening. Increasing the input plane-wave amplitude leads to higher-order $k$-gap soliton formation (Fig. 3d). Conversely, when the system is initialized with a localized wavepacket (e.g., $A_0 \text{sech}\, x$) at a small amplitude ($A_0 = 0.0005/\sqrt{2}$), an event soliton forms, fully confined in both space and time (Fig. 3b). A higher input amplitude yields higher-order event solitons (Fig. 3e), and when the amplitude reaches exactly $1/\sqrt{2}$, the event soliton transitions to a Bragg soliton, spatially confined and unaffected by $k$-gap amplification (Fig. 3c). Further increases in amplitude generate higher-order Bragg solitons (Fig. 3f). Notably, unlike the balance required for the formation of $k$-gap soliton, Bragg soliton requires a balance between nonlinearity and $\omega$-gap opening, while the event soliton results from a three-way balance.

Figure 3i illustrates the transition among soliton formations, revealing a rich conversion pathway: from event soliton to higher-order event solitons, followed by Bragg soliton, high-order Bragg solitons, and back to higher-order event solitons at even higher input amplitudes. Adjusting the width of the input seed introduces additional complexity, enabling new spatiotemporal patterns such as imperfect event solitons emerging from high-order $k$-gap solitons, as depicted in Fig. 3g. Furthermore, the superluminal characteristics inherent to the $k$-gap are evident in Fig. 3h, highlighting the diverse nonlinear phenomena accessible in STCs. Overall, these findings elucidate how the interplay of initial amplitude, spatial localization, and wavepacket width governs the emergence, instablities, and transitions of spatiotemporal pattern formations.

**Further discussions.** Unlike Peregrine soliton [35–38], a spatiotemporal localization that emerges from a weak fluctuation on a nonzero background, the event soliton does not require a continuous background. However, both soliton types can exhibit high amplitudes in spacetime, seemingly materializing abruptly and vanishing without a trace. Compared to the light bullet, a three-dimensional spatiotemporally localized electromagnetic wave proposed in the 1990s [39,40] that conserves energy during propagation as a light pulse, the event soliton manifests as a transient event that cannot be tracked along a continuous timeline.

While photonic spacetime crystals (STCs) have been extensively studied from various perspectives, their practical realization remains challenging. At radio frequencies, momentum bandgap has been experimentally observed using dynamic transmission lines [41,42]. At optical frequencies, creating photonic time crystals requires high-intensity pump beams to induce periodic refractive-index changes [43], a process constrained by small nonlinear susceptibilities and the impracticality of generating pulse trains with extremely high repetition rates. Epsilon-near-zero (ENZ) materials [44–47] may overcome these limitations in the foreseeable future. Meanwhile, on two-dimensional metamaterial platforms, modulating surface boundaries can simultaneously amplify both surface and free-space modes [48], suggesting that such spacetime metamaterials could demonstrate momentum bandgap features or the $k$-gap amplification. Recently, a topological event was reported in time-varying metamaterial [49]. For realizing event solitons, modulating surface microwaves using electronic components appears to be a feasible approach within metamaterial STCs.

**Conclusion.** In this work, we introduce the concept of an event soliton in nonlinear photonic spacetime crystals, characterized by spatial confinement and temporal transience, reminiscent of an "event" in special relativity. Crucially, we scrutinized how these solitons emerge from mixed energy-momentum gaps, thereby integrating the previously distinct frameworks of Bragg and $k$-gap soliton formation. Moreover, the existence of an event soliton implies a broader range of spatiotemporal instabilities and pattern formations in systems with spacetime periodicity. While our findings open new avenues for research, several aspects remain unexplored, including higher-order event solitons, multi-soliton interactions and collision dynamics. At this stage, we believe our work lays the groundwork for elucidating the physics of $\omega k$-gaps, and stimulates further research in time-varying photonics, spacetime metamaterials, and related fields.

**Acknowledgements**

We thanks the discussions with Prof. Dmitry Skryabin. Y.P. acknowledges the support of the NSFC (No. 2023X0201-417-03) and the fund of the ShanghaiTech University (Start-up funding).

The authors declare no competing financial interests.

Correspondence and requests for materials should be addressed to Y.P.

(yiming.pan@shanghaitech.edu.cn), and Z. Fan (zhiwei.fan@newcastle.ac.uk).

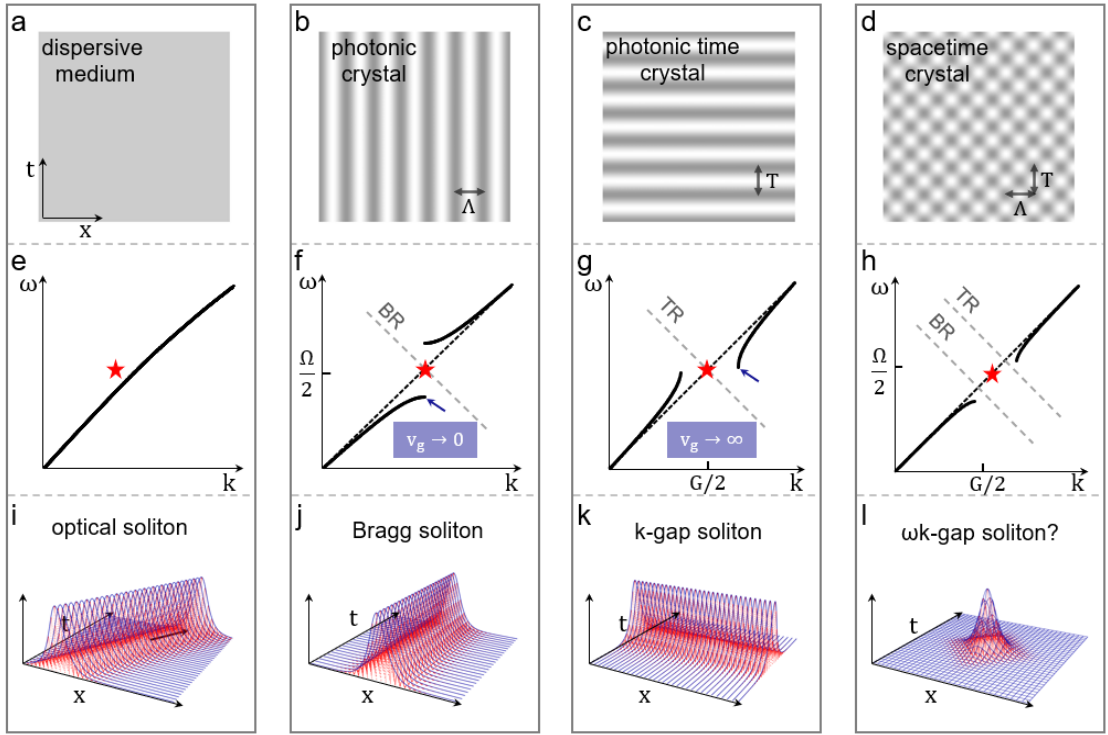

**Figure 1.** Schematic comparison of diverse photonic media, band structures and soliton formations. (a) Dispersive medium. (b) Photonic crystals with spatial modulation. (c) Photonic time crystals with temporal modulation (d) Spacetime crystals with spatiotemporal modulation. Corresponding band structures: (e) Ordinary dispersion, (f) Energy $\omega$-gapped band due to Bragg reflection (BR), (g) Momentum $k$-gapped band due to time reflection (TR), (h) Mixed $\omega k$-gapped band due to both reflections. Kerr nonlinearity is introduced to form gap solitons: (i) Traveling soliton, (j) Bragg ($\omega$-gap) soliton at zero group velocity, (k) Superluminal $k$-gap soliton at infinite group velocity, and (l) Event-like $\omega k$-gap soliton.

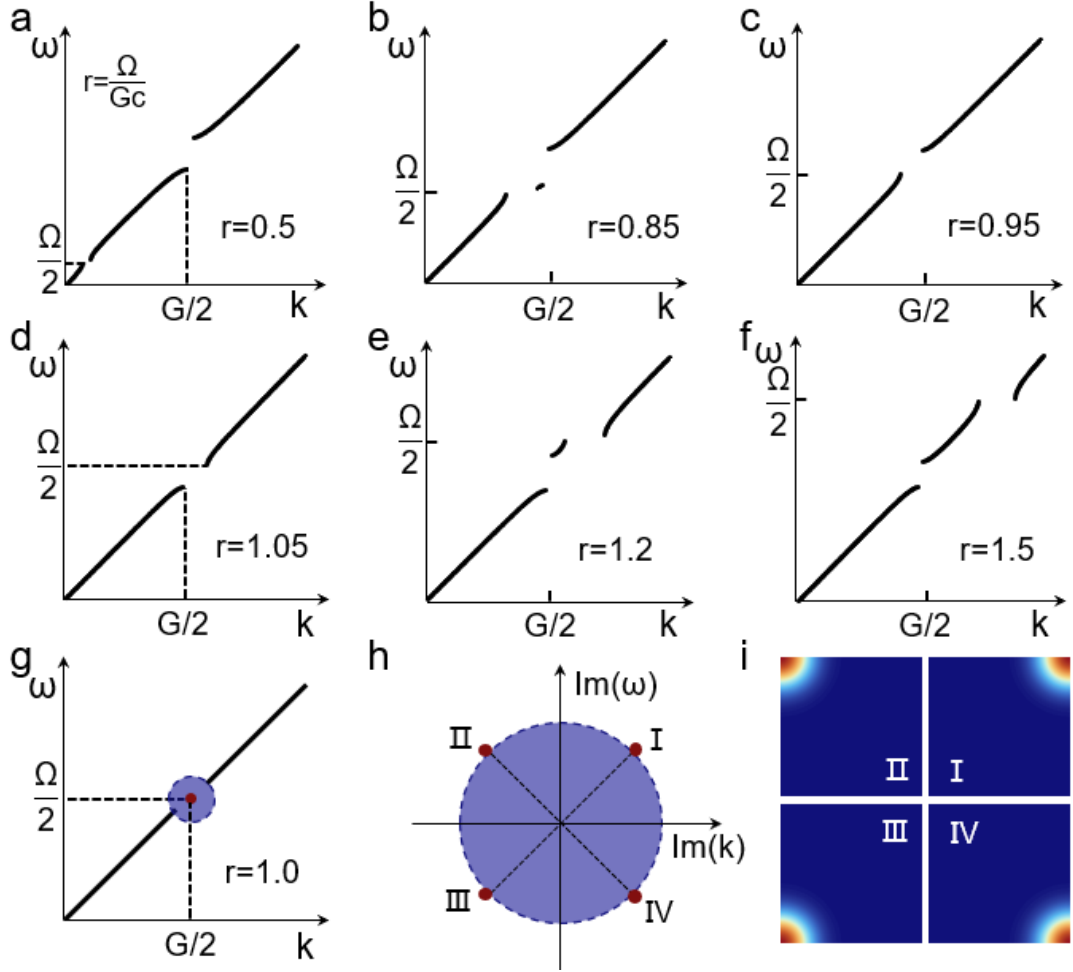

**Figure 2.** Band structure engineering and $\omega k$-gapped modes of a photonic spacetime crystal. We tune the modulation ratio $r = \Omega/Gc_r$. (a-f). As r increases, the $k$-gap progressively approaches the energy gap, forming a mixed energy-momentum gap before separating again at higher ratios. In (g) when r=1, the light-core modulation causes all $\omega k$-gapped modes to collapse to a fixed frequency and wavevector ($\Omega/2, G/2$), while its imaginary parts build up a dissipation relation in (h). The spatiotemporal patterns of these gapped modes are plotted in (i), corresponding to red points in (h).

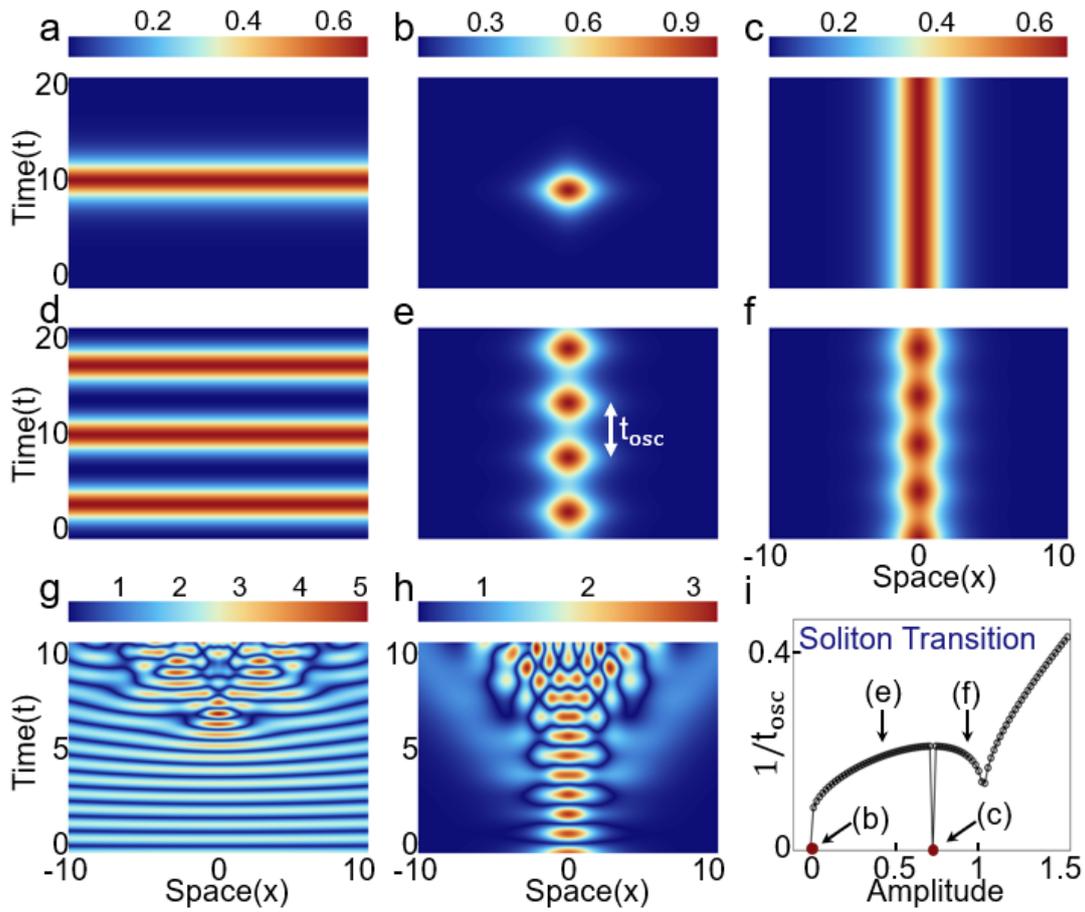

**Figure 3.** Numerical analysis of spatiotemporal pattern formations in nonlinear STCs. Evolution of fundamental gap solitons: (a) $k$-gap, (b) event, and (c) Bragg solitons. Multiple and high-order soliton generations: (d) multiple $k$-gap, (e) multiple event, and (f) high-order Bragg solitons. (g, h). Emergence of event solitons from imperfect initial conditions. (i). Transition between event and Bragg solitons by increasing the input amplitude, illustrating a conversion pathway from event, high-order Bragg to high-order event solitons.

# Supplementary Information

# Event Soliton Formation in Mixed Energy-Momentum Gaps of Nonlinear Spacetime Crystals

Liang Zhang[1], Zhiwei Fan[2], Yiming Pan[1]

[1]School of Physical Science and Technology and Center for Transformative Science, ShanghaiTech University, Shanghai 200031, China

[2]School of Engineering, Newcastle University, Newcastle upon Tyne NE1 7RU, UK

## 1. Nonlinear Photonic STCs in terms of dielectric displacement (***D***)

In lossless non-magnetic dielectric homogeneous isotropic media, Maxwell's equations are

$$\nabla \cdot E = \frac{\rho}{\epsilon_0},$$

$$\nabla \cdot B = 0,$$

$$\nabla \times E = -\frac{\partial B}{\partial t},$$

$$\nabla \times B = \mu_0 J + \mu_0 \epsilon_0 \frac{\partial E}{\partial t}, \tag{S1}$$

under time-varying permittivity, the displacement vectors $B$, $D$ remains continuous, while the fields $H$, $E$ may not necessarily continuous. This motivates using $B$ and $D$ as fundamental fields, and expressing $E$ in terms of $D$ in the following treatment of nonlinear optics. Conventionally, the dielectric displacement is expressed as a nonlinear series

$$D^i = \epsilon_0 \epsilon_1 E^i + \epsilon_0 \chi_2^{ijk} E^j E^k + \epsilon_0 \chi_3^{ijkl} E^j E^k E^l + \cdots, \tag{S2}$$

in this work, the linear part of the dielectric constant $\epsilon_1(x,t)$ is modulated periodically in space and time simultaneously. The explicit form of $\epsilon_1(x,t)$ will be provided later. The electric field E can be expressed in terms of the displacement D as

$$E^i = \frac{D^i}{\epsilon_0 \epsilon_1} - \frac{\Gamma_2^{ijk}}{\epsilon_0} D^j D^k - \frac{\Gamma_3^{ijkl}}{\epsilon_0} D^j D^k D^l - \cdots, \tag{S3}$$

with the approximated for $D^i$ reciprocal nonlinear coefficients

$$\Gamma_2^{ijk} = \frac{\chi_2^{ijk}}{\epsilon_0 \epsilon_1^3},$$

$$\Gamma_3^{ijkl} = \frac{\chi_3^{ijkl}}{\epsilon_0 \epsilon_1^4} - 2\frac{\chi_2^{ijm} \chi_2^{mkl}}{\epsilon_0 \epsilon_1^5}. \tag{S4}$$

The corresponding electromagnetic energy density ($\mathcal{H}$) is

$$\mathcal{H} = \int H \cdot dB + \int E \cdot dD = \frac{B^2}{2\mu_0} + \frac{D^2}{2\epsilon_0\epsilon_1} - \frac{\Gamma_2 D^3}{3\epsilon_0} - \frac{\Gamma_3 D^4}{4\epsilon_0} - \cdots, \quad (S5)$$

the total Hamiltonian is $H_{tot} = \int \mathcal{H} dr$. There is another reason for such a treatment addressed by Sipe, et al. [50]. The Faraday's law in (S1) is found to be violated if we take $B$, $E$ as the basis, and we check the governing dynamics $\frac{\partial B}{\partial t} = \frac{1}{i\hbar}[B, H_{tot}] = \frac{1}{i\hbar}[B, H_L + H_{NL}] = -\nabla \times E$ +other terms, where the Hamiltonian is given by $H_{tot} = \frac{1}{2\mu_0}\int B^2 dr + \frac{\epsilon_0}{2}\int \epsilon_1 E^2 dr = H_L + H_{NL}$. The other terms, including nonlinear coefficients, have incorrect signs and amplitudes. Instead, with the basis of $B$, $D$, the Maxwell equations (S1) can be recovered using the Heisenberg equations of motion, i.e., $\frac{\partial B}{\partial t} = \frac{1}{i\hbar}[B, H_{tot}], \frac{\partial D}{\partial t} = \frac{1}{i\hbar}[D, H_{tot}]$ [51].

For centrosymmetric materials, the second-order nonlinearity is negligible, i.e., $\chi_2^{ijk} = 0, \Gamma_2^{ijk} = 0$. Thus, the Hamiltonian (up to the third-order term) is reduced to

$$\mathcal{H} = \frac{B^2}{2\mu_0} + \frac{D^2}{2\epsilon_0\epsilon_1} - \frac{\Gamma_3 D^4}{4\epsilon_0}, \quad (S6)$$

where the linear dielectric constant $\epsilon_1(x,t)$ is spacetime-periodically modulated. Henceforth, we assume a simple modulation form

$$\epsilon_1(x,t) = \epsilon_r \tilde{\epsilon}(x)\tilde{\epsilon}(t) = \epsilon_r(1 + \delta_1 \cos\Omega t + \delta_2 \cos Gx), \quad (S7)$$

with the temporal modulation frequency $\Omega = \frac{2\pi}{T}$ and the spatial modulation frequency $G = \frac{2\pi}{\Lambda}$, with $\delta_{1,2} \ll 1$ being small dimensionless quantities. Correspondingly, the third-order nonlinear coefficient can be simplified to

$$\Gamma_3^{ijkl} = \frac{\chi_3^{ijkl}}{\epsilon_0 \epsilon_1^4} - 2\frac{\chi_2^{ijm}\chi_2^{mkl}}{\epsilon_0 \epsilon_1^5} \approx \frac{\chi_3}{\epsilon_0 \epsilon_1^4}. \quad (S8)$$

Leading to the Hamiltonian of nonlinear STCs

$$\mathcal{H} = \left(\frac{B^2}{2\mu_0} + \frac{D^2}{2\epsilon_0\epsilon_r}\right) - \left(\frac{\delta_1 \cos\Omega t + \delta_2 \cos Gx}{2\epsilon_0\epsilon_r}\right)D^2 - \left(\frac{\chi_3}{4\epsilon_0\epsilon_1^4}\right)D^4, \quad (S9)$$

For clarity, we summarize the governing equations (S1) discussed above, and the Maxwell equations are given by

$$\frac{\partial B}{\partial t} = -\nabla \times E,$$

$$\frac{\partial D}{\partial t} = \frac{1}{\mu_0}\nabla \times B, \qquad (S10)$$

where the electric field E is expanded in terms of the dielectric displacement D (S3), in a compact form, given by

$$E = \frac{D}{\epsilon_0 \epsilon_1} - \frac{\Gamma_3 D^3}{\epsilon_0} = \frac{D}{\epsilon_0 \epsilon_1} - \frac{\chi_3 D^3}{\epsilon_0^2 \epsilon_1^4}, \qquad (S11)$$

here, only the Kerr nonlinearity is considered, $\Gamma_3 = \frac{\chi_3}{\epsilon_0 \epsilon_1^4}$. Substituting Eq. (S11) into Eq. (S10) and eliminating the magnetic field B yields the second-order differential equations for D,

$$\frac{\partial^2 D}{\partial t^2} = -\frac{1}{\mu_0}\nabla \times (\nabla \times E) = \frac{1}{\mu_0}\frac{\partial^2}{\partial x^2}\left(\frac{D}{\epsilon_0 \epsilon_1} - \frac{\chi_3 D^3}{\epsilon_0^2 \epsilon_1^4}\right). \qquad (S12)$$

This differs from the traditional form of the wave equation for E. Noting that the parameters $\delta_{1,2}$ are small quantities, we ignore the product term $(\delta_1 \delta_2)$. We redefine the displacement as a rescaled quantity

$$\tilde{E} = \frac{D}{\tilde{\epsilon}(x)}, \qquad (S13)$$

thus, the equation can be rewritten as

$$\frac{\tilde{\epsilon}(x)}{c_r^2}\frac{\partial^2 \tilde{E}}{\partial t^2} = \frac{1}{\tilde{\epsilon}(t)}\frac{\partial^2 \tilde{E}}{\partial x^2} - \beta|\tilde{E}|^2\frac{\partial^2 \tilde{E}}{\partial x^2}, \qquad (S14)$$

with the effective speed of light in material is $c_r = 1/\sqrt{\mu_0 \epsilon_0 \epsilon_r} = c/n_0$ and the rescaled nonlinear Kerr coefficient $\beta = 3\chi_3/\epsilon_0 \epsilon_r^3$. Note that we ignore the influence of periodic modulations in the leading Kerr nonlinear term as long as the coefficient $\chi_3$ is sufficiently small. Therefore, the above nonlinear wave equation is given by

$$\frac{(1+\delta_2 \cos Gx)}{c_r^2}\frac{\partial^2 \tilde{E}}{\partial t^2} = \frac{1}{(1+\delta_1 \cos\Omega t)}\frac{\partial^2 \tilde{E}}{\partial x^2} - \beta|\tilde{E}|^2\frac{\partial^2 \tilde{E}}{\partial x^2}, \quad (S15)$$

since $\delta_1 \ll 1$, we approximate $\frac{1}{1+\delta_1 \cos\Omega t} \approx 1 - \delta_1 \cos\Omega t$, so the above equation can be written as

$$\frac{(1+\delta_2 \cos Gx)}{c_r^2}\frac{\partial^2 \tilde{E}}{\partial t^2} = (1-\delta_1 \cos\Omega t)\frac{\partial^2 \tilde{E}}{\partial x^2} - \beta|\tilde{E}|^2\frac{\partial^2 \tilde{E}}{\partial x^2}. \quad (S16)$$

## 2. The nonlinear coupled-mode equations

In this section, we derive the dispersion-dissipation relation, identify the first four photonic bands without the nonlinear term, and thus obtain the $\omega k$-gap soliton by adding the nonlinearity. Using coupled-mode theory for the slowly varying envelope, we express the dielectric field as a superposition of suitably modulated forward and backward waves,

$$\tilde{E}(x,t) = A_f e^{i\frac{G}{2}x - i\frac{\Omega}{2}t} + A_b e^{-i\frac{G}{2}x - i\frac{\Omega}{2}t} + A_f^* e^{-i\frac{G}{2}x + i\frac{\Omega}{2}t} + A_b^* e^{i\frac{G}{2}x + i\frac{\Omega}{2}t}, \quad (S17)$$

due to the dispersion relation, $\Omega \approx Gc_r$. $A_f e^{i\frac{G}{2}x - i\frac{\Omega}{2}t}$ represents the initial forward propagation, $A_b e^{-i\frac{G}{2}x - i\frac{\Omega}{2}t}$ represents the spatial reflection (backward propagation), $A_f^* e^{-i\frac{G}{2}x + i\frac{\Omega}{2}t}$ represents the simultaneous reflection in time and space (forward propagation), $A_b^* e^{i\frac{G}{2}x + i\frac{\Omega}{2}t}$ represents the time-reflected (backward propagation). Substituting this ansatz into Eq. (S16), and applying the slowly varying envelope approximation, we obtain the following coupled-mode equations for $A_f, A_b, A_f^*$ and $A_b^*$:

$$iG\frac{\partial A_f}{\partial x} + i\frac{\Omega}{c_r^2}\frac{\partial A_f}{\partial t} + \left(\frac{\Omega^2}{4c_r^2} - \frac{G^2}{4}\right)A_f + \frac{\delta_1 G^2}{8}A_b^* + \frac{\delta_2 \Omega^2}{8c_r^2}A_b$$
$$+ \frac{3}{4}\beta G^2(|A_f|^2 + 2|A_b|^2)A_f = 0,$$

$$-iG\frac{\partial A_b}{\partial x}+i\frac{\Omega}{c_r^2}\frac{\partial A_b}{\partial t}+\left(\frac{\Omega^2}{4c_r^2}-\frac{G^2}{4}\right)A_b+\frac{\delta_1 G^2}{8}A_f^*+\frac{\delta_2\Omega^2}{8c_r^2}A_f$$

$$+\frac{3}{4}\beta G^2(2|A_f|^2+|A_b|^2)A_b=0,$$

$$-iG\frac{\partial A_f^*}{\partial x}-i\frac{\Omega}{c_r^2}\frac{\partial A_f^*}{\partial t}+\left(\frac{\Omega^2}{4c_r^2}-\frac{G^2}{4}\right)A_f^*+\frac{\delta_1 G^2}{8}A_b+\frac{\delta_2\Omega^2}{8c_r^2}A_b^*$$

$$+\frac{3}{4}\beta G^2(|A_f|^2+2|A_b|^2)A_f^*=0,$$

$$iG\frac{\partial A_b^*}{\partial x}-i\frac{\Omega}{c_r^2}\frac{\partial A_b^*}{\partial t}+\left(\frac{\Omega^2}{4c_r^2}-\frac{G^2}{4}\right)A_b^*+\frac{\delta_1 G^2}{8}A_f+\frac{\delta_2\Omega^2}{8c_r^2}A_f^*$$

$$+\frac{3}{4}\beta G^2(2|A_f|^2+|A_b|^2)A_b^*=0.$$

$$(S18)$$

The nonlinear part of the equation includes self-phase modulation and cross-phase modulation terms. Simplifying the above equation, we define $\lambda=\left(\frac{\Omega^2}{4c_r^2}-\frac{G^2}{4}\right)$, $\kappa_1=\frac{\delta_1 G^2}{8}$, $\kappa_2=\frac{\delta_2\Omega^2}{8c_r^2}$, $\gamma=-\frac{3}{4}\beta G^2$, with $\lambda\to 0$. Here, $\kappa_1,\kappa_2$ denote the strengths of Bragg reflection and time reflection, and $\gamma$ is the nonlinear coefficient. We can use the Pauli matrices to represent the above determinant

$$G\left(i\frac{\partial}{\partial x}\right)\sigma_z\otimes\tau_z\psi+\frac{\Omega}{c_r^2}\left(i\frac{\partial}{\partial t}\right)\sigma_z\otimes\tau_0\psi+\lambda\sigma_0\otimes\tau_0\psi+\kappa_1\sigma_x\otimes\tau_x\psi$$

$$+\kappa_2\sigma_0\otimes\tau_x\psi-\gamma\mathcal{L}_{\mathrm{NL}}[\psi]\psi=0, \qquad (S19)$$

where $\psi=(A_f,A_b,A_f^*,A_b^*)^T$, $\sigma_{0,x,y,z},\tau_{0,x,y,z}$ can be expressed as

$$\sigma_0=\begin{pmatrix}1&0\\0&1\end{pmatrix},\sigma_x=\begin{pmatrix}0&1\\1&0\end{pmatrix},\sigma_y=\begin{pmatrix}0&-i\\i&0\end{pmatrix},\sigma_z=\begin{pmatrix}1&0\\0&-1\end{pmatrix},$$

$$\tau_i=\sigma_i(i=0,x,y,z),\sigma_0\otimes\tau_x=\begin{pmatrix}\tau_x&0\\0&\tau_x\end{pmatrix},$$

and the Kerr nonlinear operator is defined as

$$\mathcal{L}_{\mathrm{NL}}[\psi]=|A_f|^2\sigma_0\otimes\frac{3\tau_0-\tau_z}{2}+|A_b|^2\sigma_0\otimes\frac{3\tau_0+\tau_z}{2}, \qquad (S20)$$

To analyze the photonic band with $\omega k$-gap, we consider linear solutions of the plane-wave form ($\gamma = 0$)

$$G\left(i\frac{\partial}{\partial x}\right)\sigma_z \otimes \tau_z \psi + \frac{\Omega}{c_r^2}\left(i\frac{\partial}{\partial t}\right)\sigma_z \otimes \tau_0 \psi + \lambda \sigma_0 \otimes \tau_0 \psi + \kappa_1 \sigma_x \otimes \tau_x \psi$$

$$+ \kappa_2 \sigma_0 \otimes \tau_x \psi = 0. \tag{S20}$$

Substituting the ansatz $\psi = \chi e^{iPx - iEt}$, where $\chi = (\chi_1, \chi_2, \chi_3, \chi_4)^T$, yields the matrix equation about E and P

$$\left(-PG\sigma_z \otimes \tau_z + \frac{E\Omega}{c_r^2}\sigma_z \otimes \tau_0 + \lambda \sigma_0 \otimes \tau_0 + \kappa_1 \sigma_x \otimes \tau_x + \kappa_2 \sigma_0 \otimes \tau_x\right)\chi e^{iPx - iEt} = 0, \tag{S21}$$

setting the determinant of the secular equation to zero, we can get the dispersion-dissipation relation between E and P

$$\left((PG)^2 - \left(\frac{E\Omega}{c_r^2}\right)^2 - \lambda^2 - \kappa_1^2 + \kappa_2^2\right)^2 = 4\left(\frac{E\Omega\lambda}{c_r^2}\right)^2 + 4(\lambda\kappa_1)^2, \tag{S22}$$

we define the modulation ratio $r = \Omega/Gc$ as a control parameter, so that the mismatch $\lambda = G^2(r^2 - 1)/4$. Then, we can obtain the E-P four-band images. Fig S1a. shows the dispersion relation of the initial forward mode, as captured in Fig S1b. For convenience, we use Fig S1a. to present the dispersion relation in the main text. As we can see, Fig. 2c, 2d in the main text are dispersion relations of the forward propagation wave ($A_f$) intercepted in Fig. S1c and Fig. S1d, respectively.

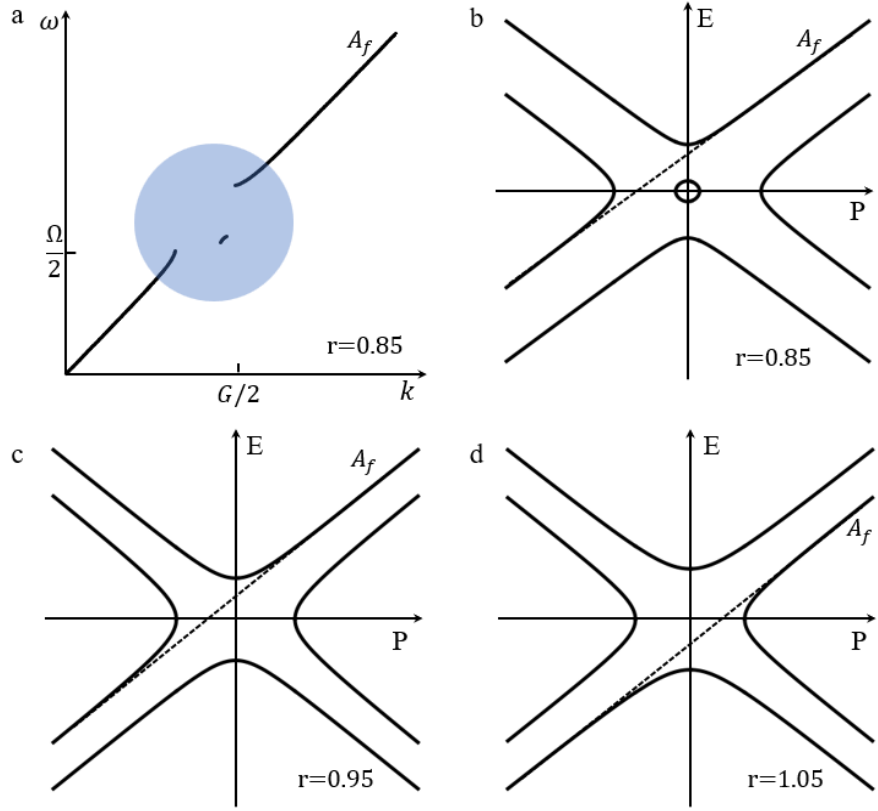

**Fig. S1:** Band structure of spacetime crystals with coupled mode theory. (a). The dispersion relation of the initial forward mode in STCs at r is equal to 0.85. (b). Exhibit how the four modes coupling within the gap region, at this time, the $\omega$-gap and $k$-gap are separated. (c). As r increases to 0.95, the two gaps approaching and merging to one gap:$\omega k$-gap. (d). When r increases to 1.05, $\omega k$-gap is still alive. But the initial forward mode position has changed in the dispersion relation.

3. **The Riemann surface of the dispersion-dissipation relation**

Here we introduce the Riemann surface to better show the band structure of STCs (see Fig. 2 in the main text). Figure S2 depict the dispersion-dissipation relations between E and P should be described in the complex-valued domains.

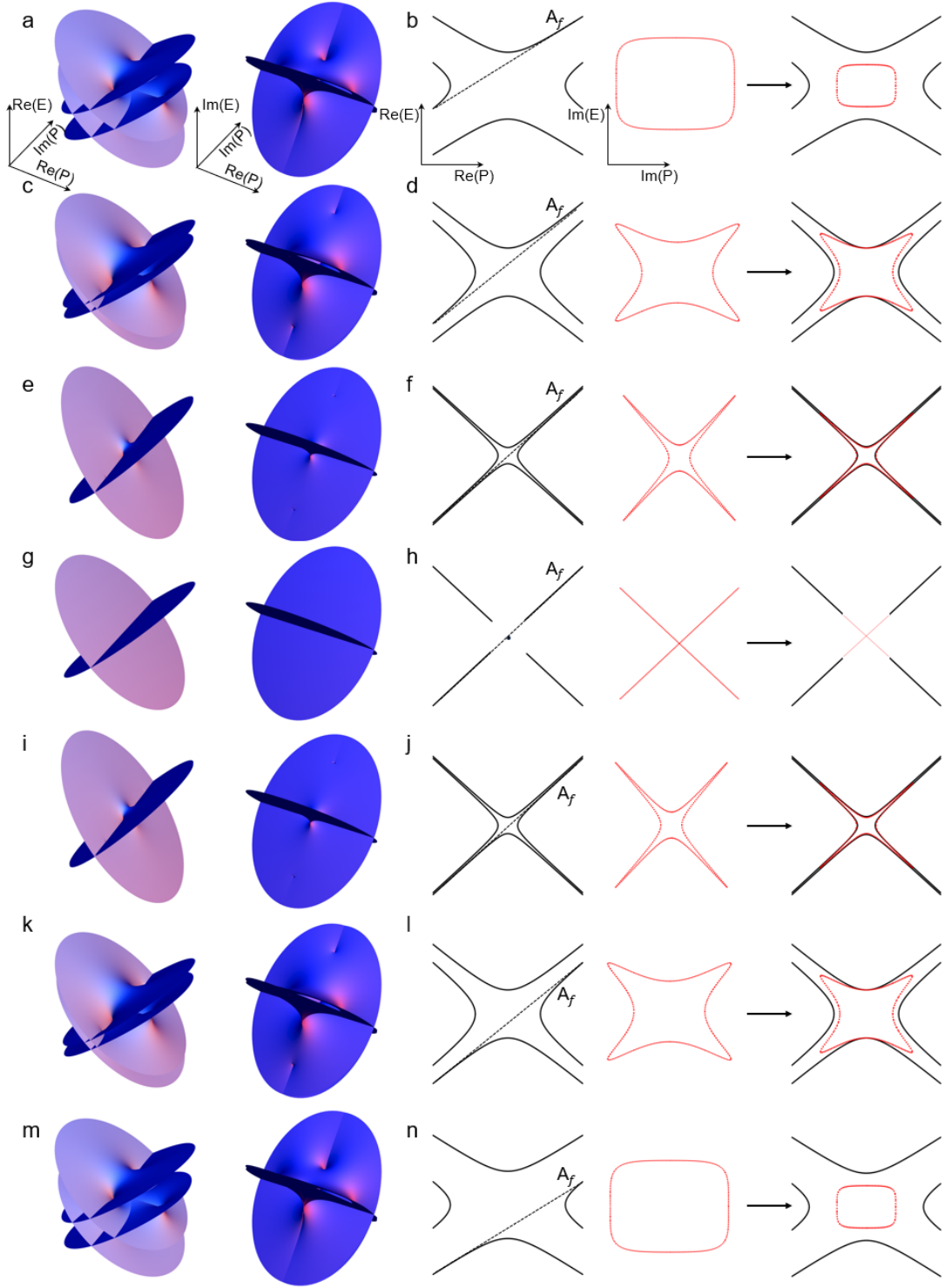

**Fig. S2:** The Riemann surfaces of fully gapped dispersion-dissipation relation of photonic STCs. At the ratio $r = 0.95$, (a) Real part of the energy Re(E) as a function of Re(P) and Im(P), and imaginary part of the energy Im(E) as a function of Re(P) and Im(P). (b) the dispersion relation (solid curve) is cut from (a) at Im(P)=0, while the dissipation relation (red curve) is cut at Re(P)=0, and their combination. (c, d) show the real part, imaginary part, and dispersion relation, dissipation relation and their combination, respectively, at the ratios $r = 0.99$. (e, f) at the ratio $r = 0.999$. (g, h) at the ratio $r = 1$. (i, j) at the ratio $r = 1.001$. (k, l) at the ratio $r = 1.01$. (m, n) at the ratio $r = 1.05$.

## 4. The numerical calculation of the nonlinear coupled mode equations

We numerically solve the four nonlinear coupled-mode equations (Eq. S18) obtained above, and observed several intriguing phenomena that validate our theoretical conclusions based on the nonlinear wave equation (Eq. 11 in the main text). Initially, we set the temporal and spatial modulation amplitudes $\delta_1 = \delta_2 = 0.3$, and the temporal and spatial modulation frequency $\Omega = G = 20\pi$. With a localized input (e.g., $\frac{0.001}{\sqrt{2}} Sech(x)$) in the $\omega k$-gap and nonlinearity $\beta = 0.8$, we obtain intensity diagrams of the forward and backward propagation modes, both exhibiting event solitons (Fig. S3a and S3b). Next, we reduced the temporal modulation amplitude to 0.1, while keeping the spatial modulation amplitude unchanged, and adjusted the modulation frequency to $5\pi$. Using a localized input (e.g., $\frac{0.001}{\sqrt{2}} Sech(0.5x)$), we observed the formation of a $k$-gap soliton (Fig. S3c). Conversely, by maintaining the temporal modulation amplitude unchanged, reducing the spatial modulation amplitude to 0.18, and setting the modulation frequency to $10\pi$, we can obtain a high-order Bragg soliton with the input $\frac{0.001}{\sqrt{2}} Sech(x)$ (Fig. S3d).

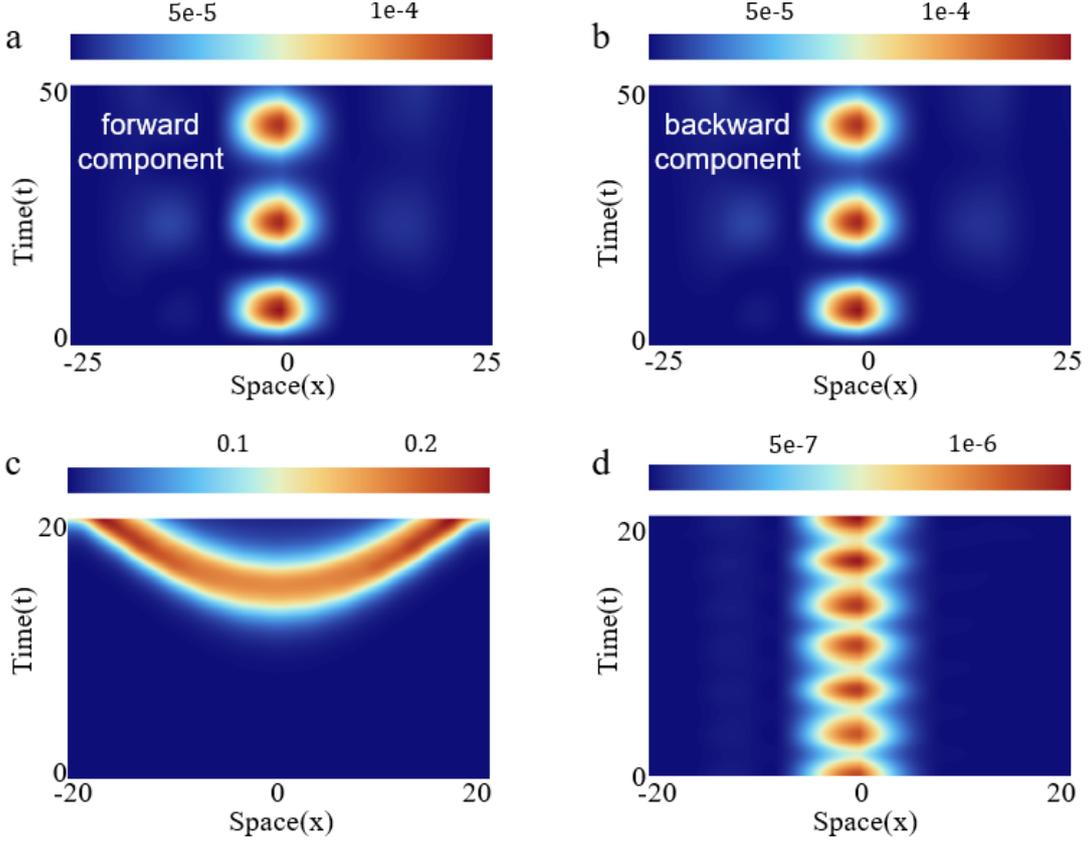

**Fig. S3:** Numerical modulation about the nonlinear coupled mode equations. (a). Event soliton of the forward modes in $\omega k$-gaps. (b). Event soliton of the backward modes in $\omega k$-gaps. (c). Decrease the spatial modulation amplitude, obtain the $k$-gap soliton. (d). Decrease the temporal modulation amplitude, we can obtain the high-order Bragg solitons.

## 5. The nonlinear Schrödinger-like equation and event solitons

Now, we consider the nonlinear effect $(\gamma \neq 0)$

$$G\left(i\frac{\partial}{\partial x}\right)\sigma_z \otimes \tau_z \psi + \frac{\Omega}{c_r^2}\left(i\frac{\partial}{\partial t}\right)\sigma_z \otimes \tau_0 \psi + \lambda \sigma_0 \otimes \tau_0 \psi + \kappa_1 \sigma_x \otimes \tau_x \psi$$
$$+\kappa_2 \sigma_0 \otimes \tau_x \psi - \gamma \mathcal{L}_{\mathrm{NL}}[\psi]\psi = 0. \tag{S23}$$

At this point, we assume a solution of the form $\psi = a(x,t)\chi e^{iP_0 x - iE_0 t}$, where $a(x,t)$ is a slowly varying amplitude. Substituting this ansatz into the above formula, yields a relation for $a(x,t)$.

For $r \to 1$, the spinors of the gapped eigenmodes are $\chi_a = (-i, i, -i, i)^T/2$, $\chi_b = (i, -i, -i, i)^T/2$, $\chi_c = (-1, -1, 1, 1)^T/2$, $\chi_d = (1,1,1,1)^T/2$, respectively. For the nonlinear term $\gamma \mathcal{L}_{NL}[\psi]\psi$, substituting one of the spinors $\chi_d$, we get

$$\mathcal{L}_{NL}[\psi] = |A_f|^2 \sigma_0 \otimes \frac{3\tau_0 - \tau_z}{2} + |A_b|^2 \sigma_0 \otimes \frac{3\tau_0 + \tau_z}{2}.$$

$$= \frac{3}{4}\gamma |a(x,t)|^2 \psi. \qquad (S24)$$

Thus, the equation for $a(x,t)$ becomes

$$\left[G\left(i\frac{\partial}{\partial x}\right)\sigma_z \otimes \tau_z + \frac{\Omega}{c_r^2}\left(i\frac{\partial}{\partial t}\right)\sigma_z \otimes \tau_0 + \lambda \sigma_0 \otimes \tau_0 + \kappa_1 \sigma_x \otimes \tau_x + \kappa_2 \sigma_0 \otimes \tau_x\right] a(x,t) \chi e^{iP_0 x - iE_0 t}$$

$$= \frac{3}{4}\gamma |a(x,t)|^2 a(x,t) \chi e^{iP_0 x - iE_0 t}.$$
$$(S25)$$

Next, we process the left side of the equation and take the Fourier transform of $a(x,t)$:

$$a(x,t) = \frac{1}{2\pi}\int dk d\omega a(k,\omega) e^{ikx - i\omega t}. \qquad (S26)$$

Substituting this into Eq. (S25), we get

$$\left[G\left(i\frac{\partial}{\partial x}\right)\sigma_z \otimes \tau_z + \frac{\Omega}{c_r^2}\left(i\frac{\partial}{\partial t}\right)\sigma_z \otimes \tau_0 + \lambda \sigma_0 \otimes \tau_0 + \kappa_1 \sigma_x \otimes \tau_x + \kappa_2 \sigma_0 \otimes \tau_x\right] a(x,t) \chi e^{iP_0 x - iE_0 t}$$
$$= \frac{1}{2\pi}\int a(k,\omega)\left[-kG\sigma_z \otimes \tau_z + \omega\frac{\Omega}{c_r^2}\sigma_z \otimes \tau_0 + \lambda \sigma_0 \otimes \tau_0 + \kappa_1 \sigma_x \otimes \tau_x + \kappa_2 \sigma_0 \otimes \tau_x\right] e^{ikx - i\omega t} \times$$
$$\chi e^{iP_0 x - iE_0 t} dk d\omega,$$
$$(S27)$$

the term in brackets can be approximated as

$$\left[-kG\sigma_z \otimes \tau_z + \omega\frac{\Omega}{c_r^2}\sigma_z \otimes \tau_0 + \lambda \sigma_0 \otimes \tau_0 + \kappa_1 \sigma_x \otimes \tau_x + \kappa_2 \sigma_0 \otimes \tau_x\right]\chi$$

$$= \left[\frac{\lambda c_r^2 + \sqrt{\omega^2 \Omega^2 + G^2 k^2 c_r^2 + c_r^4(\kappa_1^2 + \kappa_2^2) + 2\sqrt{\omega^2 k^2 G^2 \Omega^2 c_r^4 + k^2 G^2 c_r^8 \kappa_1^2 + \omega^2 \Omega^2 c_r^4 \kappa_2^2 + c_r^8 \kappa_1^2 \kappa_2^2}}}{c_r^2}\right]\chi.$$
$$(S28)$$

Now, we perform a Taylor expansion at $\omega = E_0, k = P_0$, where $E_0 \to 0, P_0 \to 0$.

$$\frac{\lambda c_r^2 + \sqrt{\omega^2\Omega^2 + G^2k^2c_r^2 + c_r^4(\kappa_1^2 + \kappa_2^2) + 2\sqrt{\omega^2k^2G^2\Omega^2c_r^4 + k^2G^2c_r^8\kappa_1^2 + \omega^2\Omega^2c_r^4\kappa_2^2 + c_r^8\kappa_1^2\kappa_2^2}}}{c_r^2}$$

$$= \lambda + \kappa_1 + \kappa_2 + \frac{\Omega^2}{2\kappa_1 c_r^4}(\omega - E_0)^2 + \frac{G^2}{2\kappa_2}(k - P_0)^2 + o^4(k,\omega). \tag{S29}$$

To obtain the desired solution, we need to take a small amount from the higher-order term $o^4(k,\omega)$

$$\frac{G^2\Omega^2}{4\kappa_1\kappa_2(\lambda + \kappa_1 + \kappa_2)c_r^4}(\omega - E_0)^2(k - P_0)^2, \tag{S30}$$

substituting this result into the original equation

$$\frac{1}{2\pi}\int a(k,\omega)\left[-kG\sigma_z \otimes \tau_z + \omega\frac{\Omega}{c_r^2}\sigma_z \otimes \tau_0 + \lambda\sigma_0 \otimes \tau_0 + \kappa_1\sigma_x \otimes \tau_x + \kappa_2\sigma_0 \otimes \tau_x\right]e^{ikx-i\omega t} \times$$
$$\chi e^{iP_0 x - iE_0 t}dkd\omega$$
$$= \frac{1}{2\pi}\int dkd\omega a(k,\omega)\chi e^{ikx-i\omega t}e^{iP_0 x - iE_0 t} \times$$
$$\left[\lambda + \kappa_1 + \kappa_2 + \frac{\Omega^2}{2\kappa_1 c_r^4}(\omega - E_0)^2 + \frac{G^2}{2\kappa_2}(k - P_0)^2 + \frac{G^2\Omega^2}{4\kappa_1\kappa_2(\lambda + \kappa_1 + \kappa_2)c_r^4}(\omega - E_0)^2(k - P_0)^2\right]$$
$$\tag{S31}$$

Then we perform the inverse Fourier transform on $a(k,\omega)$ to obtain: $a(k,\omega) = \frac{1}{2\pi}\int dx'\,dt'\,a(x',t')e^{-ikx'+i\omega t'}$, and substitute it into the above formula

$$= \left(\frac{1}{2\pi}\right)^2 \int dk d\omega \int dx' dt' a(x',t') e^{-ikx'+i\omega t'} \left[\lambda + \kappa_1 + \kappa_2 + \frac{\Omega^2}{2\kappa_1 c_r^4}(\omega - E_0)^2 \right.$$

$$\left. + \frac{G^2}{2\kappa_2}(k-P_0)^2 + \frac{G^2\Omega^2}{4\kappa_1\kappa_2(\lambda+\kappa_1+\kappa_2)c_r^4}(\omega-E_0)^2(k-P_0)^2\right] e^{ikx-i\omega t} \chi e^{iP_0 x - iE_0 t}$$

$$= e^{2(iP_0 x - iE_0 t)} \left(\frac{1}{2\pi}\right)^2 \int dk d\omega \int dx' dt' a(x',t') e^{-ikx'+i\omega t'} \left[\lambda + \kappa_1 + \kappa_2 \right.$$

$$+ \frac{\Omega^2}{2\kappa_1 c_r^4}(\omega - E_0)^2 + \frac{G^2}{2\kappa_2}(k-P_0)^2$$

$$\left. + \frac{G^2\Omega^2}{4\kappa_1\kappa_2(\lambda+\kappa_1+\kappa_2)c_r^4}(\omega-E_0)^2(k-P_0)^2 \right] e^{i(k-P_0)x - i(\omega-E_0)t}\chi$$

$$= e^{2(iP_0 x - iE_0 t)}\left[\lambda + \kappa_1 + \kappa_2 - \frac{\Omega^2}{2\kappa_1 c_r^4}\frac{\partial^2}{\partial t^2} - \frac{G^2}{2\kappa_2}\frac{\partial^2}{\partial x^2} \right.$$

$$\left. + \frac{G^2\Omega^2}{4\kappa_1\kappa_2(\lambda+\kappa_1+\kappa_2)c_r^4}\frac{\partial^4}{\partial t^2 \partial x^2}\right]\left(\frac{1}{2\pi}\right)^2 \int dk d\omega \int dx' dt' a(x',t')e^{-ikx'+i\omega t'} e^{i(k-P_0)x - i(\omega-E_0)t}\chi$$

$$= e^{(iP_0 x - iE_0 t)}\left[\lambda + \kappa_1 + \kappa_2 - \frac{\Omega^2}{2\kappa_1 c_r^4}\frac{\partial^2}{\partial t^2} - \frac{G^2}{2\kappa_2}\frac{\partial^2}{\partial x^2} \right.$$

$$\left. + \frac{G^2\Omega^2}{4\kappa_1\kappa_2(\lambda+\kappa_1+\kappa_2)c_r^4}\frac{\partial^4}{\partial t^2 \partial x^2}\right] a(x,t)\chi.$$

Equating this to the nonlinear term on the right-hand side

$$e^{(iP_0 x - iE_0 t)}\left[\lambda + \kappa_1 + \kappa_2 - \frac{\Omega^2}{2\kappa_1 c_r^4}\frac{\partial^2}{\partial t^2} - \frac{G^2}{2\kappa_2}\frac{\partial^2}{\partial x^2} + \frac{G^2\Omega^2}{4\kappa_1\kappa_2(\lambda+\kappa_1+\kappa_2)c_r^4}\frac{\partial^4}{\partial t^2 \partial x^2}\right] a(x,t)\chi$$

$$= \frac{3}{4}\gamma |a(x,t)|^2 a(x,t) \chi e^{iP_0 x - iE_0 t}, \qquad (S32)$$

we can get a partial differential equation for $a(x,t)$

$$\left[\lambda + \kappa_1 + \kappa_2 - \frac{\Omega^2}{2\kappa_1 c_r^4}\frac{\partial^2}{\partial t^2} - \frac{G^2}{2\kappa_2}\frac{\partial^2}{\partial x^2} + \frac{G^2\Omega^2}{4\kappa_1\kappa_2(\lambda+\kappa_1+\kappa_2)c_r^4}\frac{\partial^4}{\partial t^2 \partial x^2}\right] a(x,t)$$

$$- \frac{3}{4}\gamma |a(x,t)|^2 a(x,t) = 0. \qquad (S33)$$

Simplifying further, we obtain

$$\left[1 - \frac{\Omega^2}{2\kappa_1 c_r^4(\lambda+\kappa_1+\kappa_2)}\frac{\partial^2}{\partial t^2}\right]\left[1 - \frac{G^2}{2\kappa_2(\lambda+\kappa_1+\kappa_2)}\frac{\partial^2}{\partial x^2}\right] a - \frac{3\gamma}{4(\lambda+\kappa_1+\kappa_2)}|a|^2 a = 0. (S34)$$

which clearly exhibits group velocity dispersion, scattering and nonlinear effects. This partial differential equation admits soliton solutions, and we propose a trial solution $a = A\, sech(Bt)\, sech(Cx)$ into the above equation. Determine the constants A, B, and C yields

$$A = \frac{4\sqrt{\lambda + \kappa_1 + \kappa_2}}{\sqrt{3\gamma}}, B = \frac{c_r^2\sqrt{2\kappa_1(\lambda + \kappa_1 + \kappa_2)}}{\Omega}, C = \frac{\sqrt{2\kappa_2(\lambda + \kappa_1 + \kappa_2)}}{G}$$

Substituting these into the trial solution, we can get

$$a = \frac{4\sqrt{\lambda + \kappa_1 + \kappa_2}}{\sqrt{3\gamma}} sech\left(\frac{c_r^2\sqrt{2\kappa_1(\lambda + \kappa_1 + \kappa_2)}}{\Omega}t\right) sech\left(\frac{\sqrt{2\kappa_2(\lambda + \kappa_1 + \kappa_2)}}{G}x\right). (S35)$$

For a special situation (r=1), where λ=0 and $\kappa_1 = \kappa_2 = \kappa$, the equation reduces to

$$\left(1 - \frac{\Omega^2}{4\kappa^2 c_r^4}\frac{\partial^2}{\partial t^2}\right)\left(1 - \frac{G^2}{4\kappa^2}\frac{\partial^2}{\partial x^2}\right)a - \frac{3\gamma}{8\kappa}|a|^2 a = 0, \qquad (S36)$$

which admits the analytical solution as we expected in the $\omega k$-gap

$$a(x,t) = \frac{4\sqrt{2\kappa}}{\sqrt{3\gamma}} sech\left(\frac{2\kappa c^2}{\Omega}t\right) sech\left(\frac{2\kappa}{G}x\right). \qquad (S37)$$

Thus, the electric field $\tilde{E}(x,t)$ can be written as

$$\tilde{E}(x,t) = a(x,t) * \frac{1}{2}\left(e^{i\frac{G}{2}x - i\frac{\Omega}{2}t} + e^{-i\frac{G}{2}x - i\frac{\Omega}{2}t} + e^{-i\frac{G}{2}x + i\frac{\Omega}{2}t} + e^{i\frac{G}{2}x + i\frac{\Omega}{2}t}\right). \qquad (S38)$$

Accordingly, the electric displacement $D(x,t)$ is given by

$$D(x,t) = \frac{8\sqrt{2\kappa}}{\sqrt{3\gamma}}(1 + \delta_2 cosGx)sech\left(\frac{2\kappa c^2 t}{\Omega}\right) sech\left(\frac{2\kappa x}{G}\right) cos\left(\frac{Gx}{2}\right) cos\left(\frac{\Omega t}{2}\right). (S39)$$

## 6. Bandgap classification as energy $\omega$-gap, momentum k-gap, and mixed $\omega k$-gap

The opening of energy gaps, momentum gaps, and mixed $\omega k$-gaps has sparked significant interest in both fundamental physics and technological applications. While energy and momentum gaps, such as those found in semiconductors and topological insulators [52], are well-explored, the mixed $\omega k$-gap is still largely uncharted territory, representing an intriguing frontier in band structure engineering. The physics of $\omega k$-gap, referred to as "UNKNOWN UNKNOWN", combines both energy and momentum domains, leading to novel effects and phenomena that challenge traditional models and offering the potential for new quantum technologies and advanced material design. The mixed $\omega k$-gap band engineering lies in its ability to create spatiotemporal patterns and other exotic states of matter. This unexplored gap promises both theoretical insights and practical breakthroughs, particularly in fields such as nonlinear time-varying media, and periodically-driven condensed matter systems, where the interplay between energy and momentum is crucial, driving further research into its foundations and applications.

| "Mind the gap!" | Energy $\omega$-gap (KNOWN KNOWN) | Momentum $k$-gap (KNOWN UNKNOWN) | Mixed $\omega k$-gap (UNKNOWN UNKNOWN) |
|---|---|---|---|
| Physics and applications | Semiconductor<br>Phase Transition<br>Topological Insulators<br>Bragg Soliton<br>…… | Parametric Amplification<br>Exceptional Point<br>Photon Pair Generation<br>$k$-gap Soliton<br>…… | Event Soliton<br>…… |

**Table S1.** The physics and applications associated with the opening of energy gap, momentum gap, and mixed gap. We describe the physics of $\omega k$-gap as "UNKNOWN UNKONWNS" in an informal way, in order to address its exploration and novelty.